# Charge-transfer effect on local lattice distortion in a HfNbTiZr high entropy alloy

Fanchao Meng[a1], Wenyan Zhang[a1], Zhukun Zhou[b], Ruixin Sheng[a], Andrew C. -P. Chuang[c], Chongchong Wu[a], Hailiang Huang[a], Xia Li[a], Shangzhou Zhang[a], Hua Zhang[a], Lilong Zhu[a], Liang Jiang[a], Peter K. Liaw[d], Shuying Chen[a*] & Yang Tong[a*]

[a]Institute for Advanced Studies in Precision Materials, Yantai University, Yantai, Shandong 264005, China

[b]Institute of Laser Intelligent Manufacturing and Precision Processing, School of Mechanical Engineering, Guangxi University, Nanning, Guangxi 530004, China

[c]Argonne National Laboratory, Lemont, IL 60439, USA

[d]Department of Materials Science and Engineering, The University of Tennessee, Knoxville, TN 37996, USA

[1]F. Meng and W. Zhang contribute equally to this work.

*Corresponding author: sychen@ytu.edu.cn, yt1@ytu.edu.cn.

**Abstract**

It is often assumed that atoms are hard spheres in the estimation of local lattice distortion (LLD) in high-entropy alloys (HEAs). However, our study demonstrates that the hard sphere model misses the key effect, charge transfer among atoms with different electronegativities, in the understanding of the stabilization of severely-distorted HEAs. Through the characterization and simulations of the local structure of the HfNbTiZr HEA, we found that the charge transfer effect competes with LLD to significantly reduce the average atomic-size mismatch. Our finding may form the basis for the design of severely distorted, but stable HEAs.

The rapid development of high entropy alloys (HEAs), solid-solution alloys consisting of multiple components at equiatomic or near-equiatomic concentration[1, 2], has made discoveries of unusual edge dislocation-controlled strengthening mechanism and outstanding radiation resistance particularly in the body-centered cubic (BCC) refractory HEAs[3-5]. Compared with the most studied face-centered cubic (FCC) HEAs composed of 3$d$ transition metals only[6-8], what makes refractory HEAs unique is the diversity of local atomic environment caused by the large difference of electronegativity and atomic size among the constitutes including 3$d$, 4$d$, 5$d$ elements[9-11]. For example, the local lattice distortion (LLD), considered as one of the core effects in HEAs, has been largely appreciated by both experimental and theoretical studies of BCC refractory HEAs[11-14]. In particular, the Zr/Hf-containing refractory HEAs exhibit severe LLD with the atomic displacements meeting the Lindemann melting criterion[11], leading to a significant improvement of yield strength but without a sacrifice of ductility[15].

However, less attention has been paid to the importance of local chemistry (i.e. electronegativity) difference in the refractory HEAs. Recent studies found that BCC HEAs with negligible LLD also show extraordinarily high yield strengths by greatly retarding the motion of edge dislocations[3, 4, 16]. These BCC HEAs, however, have a typical large difference of electronegativity among constitute components. Thus, mobile dislocations inevitably experience the charge transfer effect (CTE) due to the change of local environments, which is often ignored because of the difficulty in its characterization. But the inclusion of CTE is critical for the understanding of BCC HEAs' unusual behaviors.

As a demonstration of the equal importance of CTE and LLD, we examined the influence of CTE on LLD in the most studied BCC refractory HEA, HfNbTiZr, through a combination of local structure characterization and different simulation methods in the present work. Our study clearly revealed that the CTE plays a critical role in the stabilization of the HfNbTiZr HEA with a severe LLD. Moreover, our finding of local environment-dependent CTE highlights another tuning parameter for the design of high-performance HEAs.

The HfNbTiZr sample was prepared by arc melting Hf, Nb, Ti, and Zr metals ($\geq$ 99.5 wt% purity). The arc-melted button was flipped and remelted five times before dropping cast into a rectangular copper mold. The drop-cast ingot was then sealed into a vacuum quartz tube and homogenized at 1,473 K for 24 h, followed by water quenching. Some powder was ground from the homogenized specimen for the X-ray total scattering experiment. The total scattering measurements were conducted at the 11-ID-B beamline of the Advanced Photon Source with an X-ray energy of 58.6 keV ($\lambda$ = 0.2116 Å). The PDFgetX3[17] was used to obtain an atomic pair distribution function (PDF) by a Fourier transformation of the measured total scattering structure function, S(Q),

$$G(r) = \frac{2}{\pi}\int_0^Q [S(Q) - 1]\sin(Qr)\, dQ.$$

Here, $Q$ is the scattering vector and $r$ the real-space interatomic distance. The $Q$ range for the Fourier transformation is 20 Å$^{-1}$. To find the atomic configuration solution for the HfNbTiZr HEA, we used PDFgui[18] and RMCprofile[19] software to fit the measured PDF.

To examine the LLD in HfNbTiZr, a non-distorted BCC structure model with randomly distributed Hf, Nb, Ti and Zr atoms shown in Fig. 1a was fitted to the experimental PDF over an interatomic distance range of 30 Å. Here, the non-distorted supercell containing 250 atoms was constructed, utilizing the special quasi-random structure (SQS) approach[20]. Fig. 1b compares the experimental and fitted PDFs by showing their difference curve. The difference curve indicates that the first and second PDF peaks cannot be fitted by the non-distorted SQS model while beyond these two PDF peaks, the experimental PDF agrees well with the fitted one, demonstrating the deviation of the local structure from the average structure. Details about the deviation can be seen from a close view of the local structure fitting, as shown in Fig. 1c. It can be seen that the first and second peaks of the experimental PDF shift towards each other to form a broad atomic shell. In other words, the average inter-atomic distance among the first nearest neighboring atoms (1NNA) increases, while the second nearest neighboring atoms (2NNA) shorten their average distance. These features prove the existence of a large LLD in the HfNbTiZr HEA.

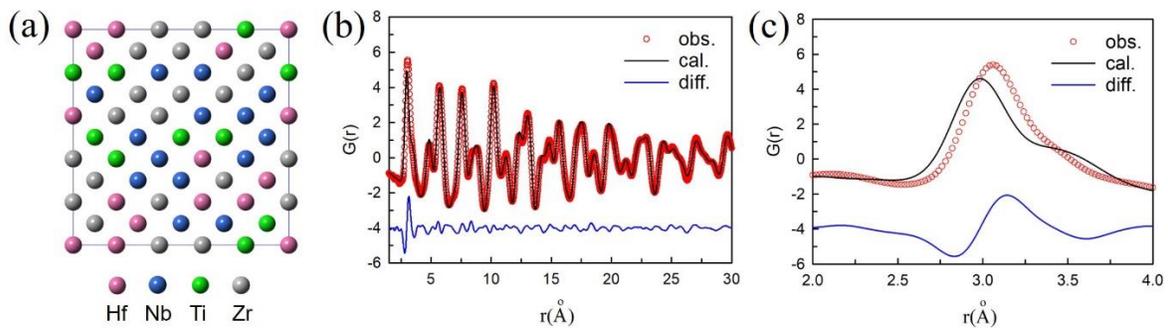

**Figure 1.** (a) SQS supercell with an ideal BCC crystal structure. (b) ideal structure fit to the PDF of HfNbTiZr. (c) Close view of the fitting of the first and second PDF peaks in (b).

To explore the possible structural configuration in HfNbTiZr, we implemented a Reverse Monte Carlo (RMC) method, a hard sphere approach, to reproduce the experimental PDF[20]. In the RMC fitting, a large simulation box with 21,296 atoms arranged in a random fashion was constructed first (see Figure S1 in the supplementary file). The benefit of constructing such a large simulation box is to sample all possible local atomic environments, which must be considered for the compositionally complex HEAs. In the RMC fitting procedure, atoms were allowed to translate a random distance but without atom swap moves. Hence, the random-alloy configuration was maintained. The RMC structure model and its fit to the experimental PDF were shown in Fig. 2. The RMC structure model unambiguously reveals that atoms in the HfNbTiZr HEA displace from their BCC crystallographic sites. Meanwhile, the RMC fit matches the measured PDF very well over the whole $r$ range, suggesting the structure model obtained from the RMC method can reproduce both the average and local structures. Moreover, another RMC fitting by allowing atoms both translate a random distance and swap equally with each other was performed to see if chemical ordering contributes to the LLD. To quantify the chemical ordering, we used the Warren-Cowley short-range order (SRO) parameter[21],

$$\alpha_n^{ij} = 1 - \frac{P_n^{ij}}{C_j}$$

where $P_n^{ij}$ describes the probability of finding an atom of type $j$ adjacent to an atom of type $i$, the subscript $n$ refers to the $n$th atomic shell, and $C_j$ is the concentration of atom $j$. The case of an ideal random alloy has a Warren-Cowley parameter of zero. A positive value of $\alpha_n^{ij}$ represents an ordered structure, while a negative value reflects chemical segregation. Our calculation shows that the inclusion of swap move introduces some Nb-Nb segregation (Figure S2), but the fitting is not improved by the Nb-Nb segregation (Figure S3). It is apparent that the local Nb-Nb segregation has negligible influence on the severe LLD in HfNbTiZr HEA. Therefore, the severe LLD found in the HfNbTiZr HEA is purely attributed to the atomic displacements induced by size mismatch among constitues.

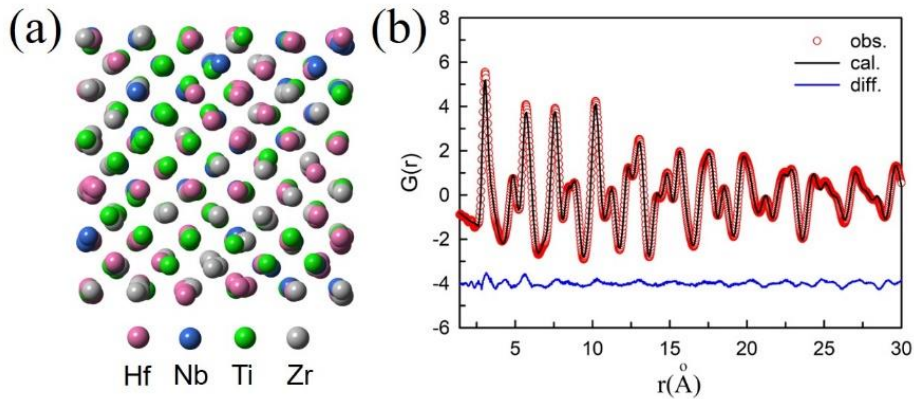

**Figure 2.** (a) Structure model obtained from the RMC fit. (b) RMC fit to the PDF of HfNbTiZr. Here, only a small portion of atoms in the RMC simulation box are shown to clearly demonstrate LLD.

The RMC modeling agrees with the experimental PDF, but the thermodynamic stability of the structure model is not considered. The introduction of atomic displacement in the RMC simulation inevitably increases the energy of a modeled system, which destabilizes the system. Using the density functional theory (DFT) method, Song et al.[14] calculated the LLD energy and the enthalpy change after the introduction of LLD into the modeled HEAs. Expectedly, they found that LLD increases the energy of HEAs, but unexpectedly the enthalpy of all studied HEAs decreases by almost the same amount when the LLD is introduced. The reduced enthalpy can easily cause the confusion that LLD can stabilize the modeled alloy systems. Instead, the enthalpy reduction is attributed to the chemical effect induced by charge transfer among atoms with different electronegativity. From the energy perspective, CTE competes with LLD to reduce the energy cost induced by LLD.

We further conducted the DFT calculation to examine the CTE on the LLD of the HfNbTiZr HEA. The DFT calculations were carried out by using the VASP code with the projector augmented wave method[22, 23]. The exchange-correlation functional was chosen as the generalized gradient approximation in the Perdew-Burke-Ernzerhof form[24]. A $2 \times 2 \times 2$ $\Gamma$-centered $k$-point mesh was chosen for the Brillouin Zone sampling. The plane-wave energy cutoff was 300 eV. The energy and force tolerance were $10^{-4}$ eV and 0.01 eV·Å$^{-1}$, respectively. A SQS supercell with 250 atoms was constructed for the DFT calculations. The DFT simulations were performed to find the optimal volume with the lowest energy and then to relax the atomic positions by fixing the supercell volume.

Figure 3a shows a relaxed random structure from the DFT calculation. Similar to the RMC simulation case, atoms in this relaxed supercell also displace from their ideal BCC crystallographic positions. Still, it can be seen that atoms in the DFT supercell are less distorted than those in the RMC supercell presented in Fig. 2a. To test whether the DFT supercell can reproduce the experimental PDF, we further fitted the relaxed supercell to the experimental data, as shown in Fig. 3b. The fit matches the observed PDF very well, including the local structure in the small $r$ range. The agreement of the DFT supercell with the experimental PDF suggests that atoms displace less when charge transfer occurs.

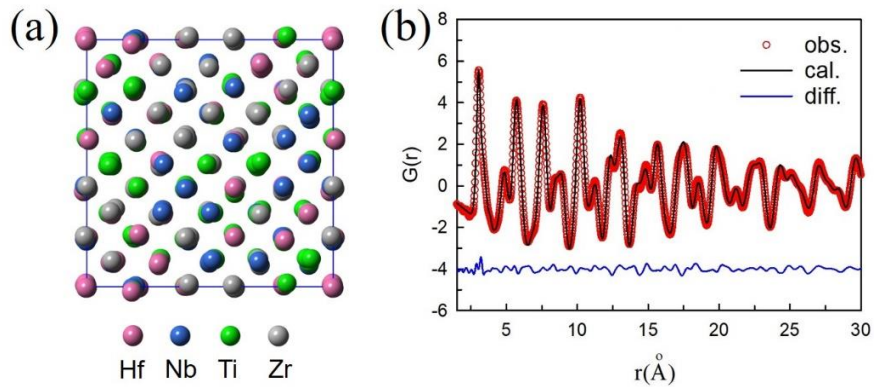

**Figure 3.** (a) Relaxed supercell obtained from DFT calculation. (b) Relaxed structure fit to the experimental PDF of HfNbTiZr.

In the RMC simulation, atoms are assumed as hard spheres without the consideration of CTE. However, when forming alloys constituent atoms can lose or gain electrons to vary sizes according to their electronegativity difference. Fig. 4a shows the charge change of $d$ ($t_{2g}$) orbitals, $\Delta\rho$, for each atom in the DFT supercell after the relaxation of atomic positions. Here, a positive value of $\Delta\rho$ indicates gaining electrons while a negative value means losing electrons. It can be seen that the majority of small Nb and Ti atoms attract electrons whereas large Hf and Zr atoms donate electrons. Therefore, there exists a charge transfer from large Hf and Zr atoms to small Nb and Ti ones, which is expected since Hf and Zr atoms have smaller electronegativity values than Nb and

Ti atoms. As a consequence, large Hf and Zr atoms decrease their sizes while the size of small Nb and Ti atoms increases. But the extent of the atomic-size change depends on the local atomic environments, which results in the fluctuation of the atomic size for each element. Taking the Zr element as an example, Figs. 4b and c show the local configurations of the largest and smallest Zr atoms, respectively. We can see that the largest Zr atom has more like-atoms, Zr and Hf, as the 1NNAs and 2NNAs, whereas the smallest Zr atom has more unlike-atoms, such as Ti and Nb, as the 1NNAs and 2NNAs, leading to a higher CTE around the smallest Zr atom. A similar trend is also found for the local configurations of the largest and smallest Hf atoms, exhibited in Figs. 4d and e.

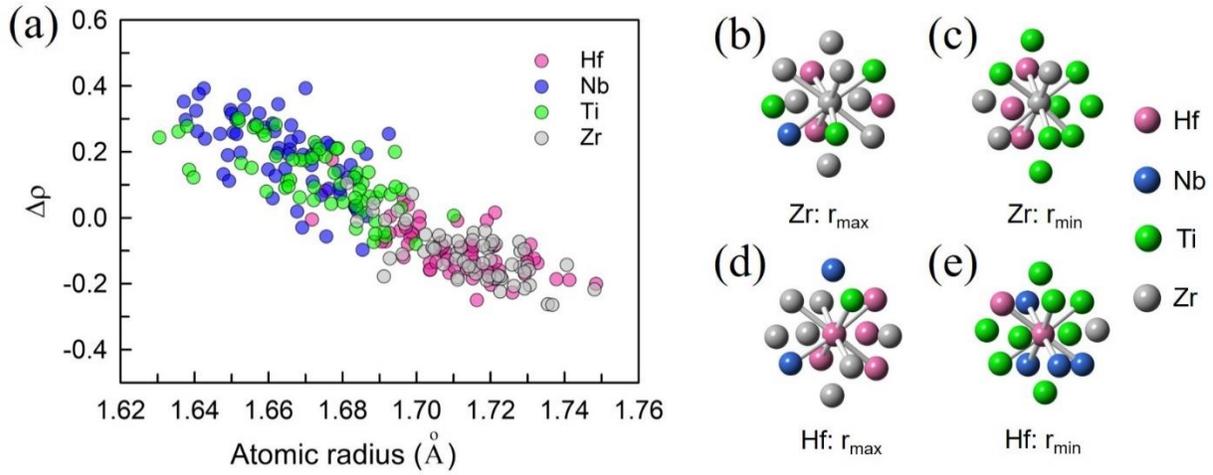

**Figure 4.** (a) Correlation between charge transfer for $d$ ($t_{2g}$) orbitals and atomic radius. (b-e) Local configurations of the largest and smallest Zr/Hf atoms.

A widely used parameter to describe the LLD of HEAs is the size-mismatch parameter

$$\delta = \sqrt{\sum_{i=1}^{N} c_i (1 - r_i / \sum_{j=1}^{N} c_j r_j)^2} \qquad (1)$$

where $N$, $c_{i,j}$, and $r_{i,j}$ are the total number of constituent elements, atomic fraction, and atomic radius of the ith or jth element, respectively[25, 26]. However, the CTE has been rarely considered in the LLD evaluation for HEAs by treating the constituent element with a fixed atomic size. Here, we quantitatively analyzed the CTE on the LLD of the HfNbTiZr HEA. Table 1 shows the CTE on the atomic radius for each element in the HfNbTiZr HEA. For the hard sphere case, the atomic radius, $r_i$, is obtained from the measured lattice constants of pure BCC metals[27] through $a^3/2 =$

¾ × π × $r_i^3$, where a is the lattice constant. For the soft sphere case, the atomic radius for each element is the average value calculated from the DFT supercell. It is obvious that the formation of the HfNbTiZr HEA alters the atomic radii of their constituent elements. Through charge transfer, the size mismatch is dramatically reduced from 7.3% to 1.4%, a decrease of ~80%. Therefore, the CTE reduces the extent of the average LLD in the HfNbTiZr HEA to stabilize its distorted structure. Meanwhile, we should realize that lattice distortion locally can still be very large, depending on the local environments especially where the like atoms, e.g. Zr and Hf, are neighboring to each other (Fig. 4b). The broad distribution of atomic radius for each element in Fig. 4 reflects the complexity of local atomic environments.

**Table 1.** Charge transfer effect on atomic radii and size mismatch for the HfNbTiZr HEA.

|  | Atomic radius (Å) | | | | δ (%) |
| --- | --- | --- | --- | --- | --- |
|  | Hf | Nb | Ti | Zr |  |
| Hard sphere | 1.78 | 1.62 | 1.63 | 1.78 | 7.3 |
| Soft sphere | 1.71 | 1.66 | 1.67 | 1.71 | 1.4 |

In summary, we investigated the LLD in the HfNbTiZr HEA through a combination of local structure characterization and simulations. Our study found that the local structure of the HfNbTiZr HEA deviates from its average structure by an expansion of the distance between the first nearest neighboring atoms and meanwhile a shrink of the distance of the second nearest neighboring atoms. This energy-unfavorable distorted local structure, however, is stabilized by the charge transfer from the large Hf and Zr atoms to the small Ti and Nb atoms. Although lattice distortion can be severe particularly in some local regions with less charge transfer, the CTE significantly reduces the average lattice distortion.

**Acknowledgement**

Y.T. acknowledges the financial support programs, Taishan Scholars Youth Expert Program, and Top Discipline in Materials Science of Shandong Province. S.C. acknowledges the financial support from the National Natural Science Foundation of China (No. 52001271). P.K. Liaw very much appreciates the supports from (1) the National Science Foundation (DMR-1611180 and

**Supplementary file**

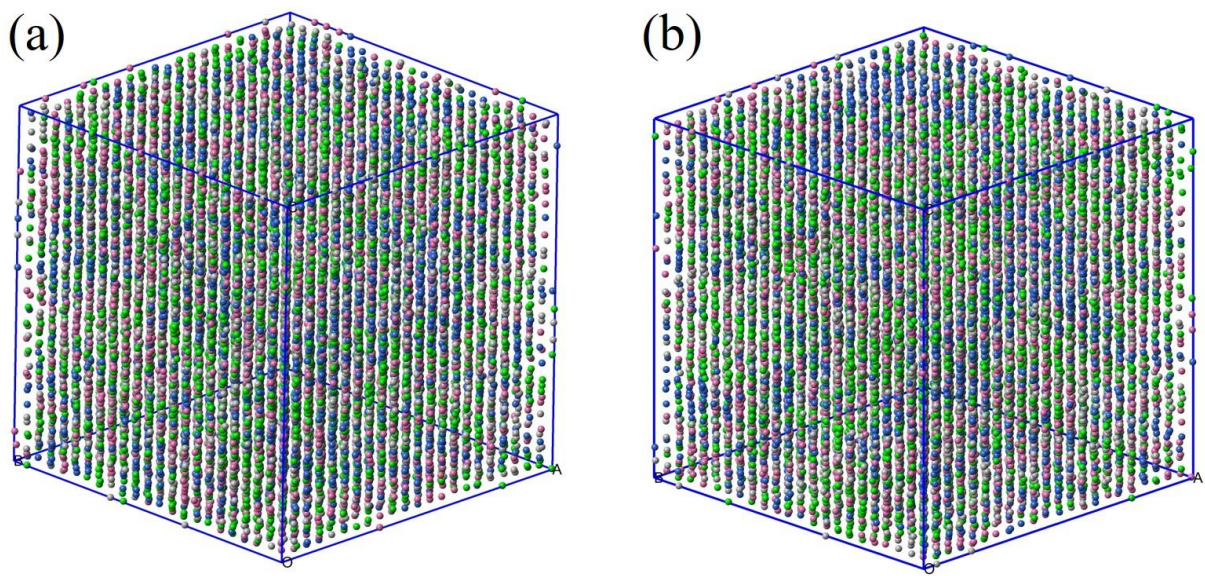

**Figure S1.** (a)RMC structure model without the atom swap move; (b) RMC structure model with the atom swap move. Here, Hf, Nb, Ti, and Zr atoms are in pink, blue, green, and grey, respectively.

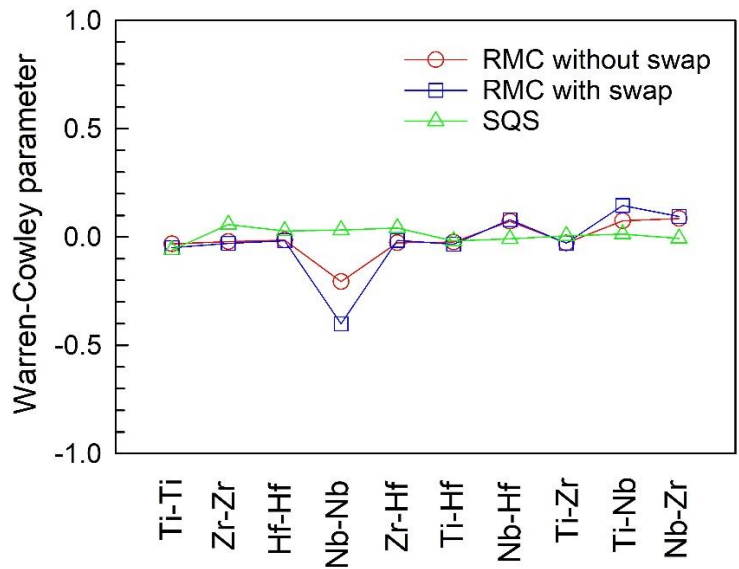

**Figure S2.** Order parameter of each atom pair for the RMC supercells and the SQS supercell used in the DFT calculation.

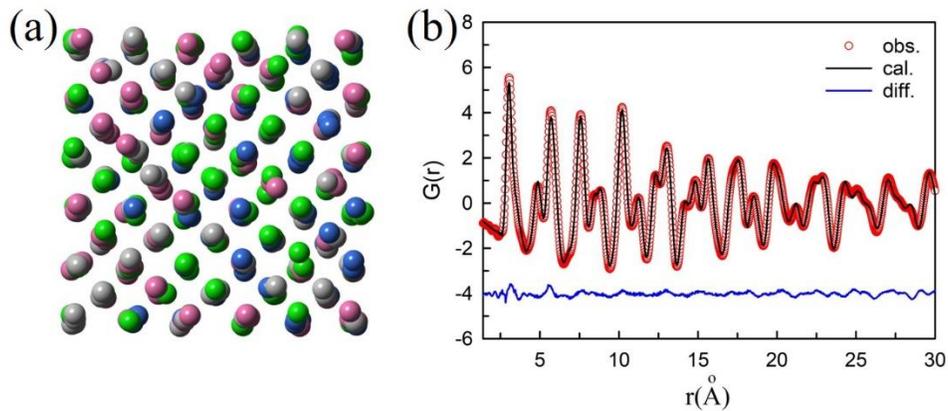

**Figure S3.** (a) RMC structure model with the atom swap move; (b) RMC fit to the experimental PDF. Here, Hf, Nb, Ti, and Zr atoms are in pink, blue, green, and grey, respectively.